\documentclass[%
 aip,
 amsmath,amssymb,
 reprint,%
]{revtex4-1}

\usepackage{graphicx}
\usepackage{dcolumn}
\usepackage{bm}

\usepackage[utf8]{inputenc}
\usepackage[T1]{fontenc}
\usepackage{mathptmx}
\usepackage{xcolor}

\newcommand*{\citen}{}
\DeclareRobustCommand*{\citen}[1]{%
  \begingroup
    \romannumeral-`\x 
    \setcitestyle{numbers}%
    \cite{#1}%
  \endgroup
}

\newenvironment{data availability}
{
  \par\vspace{\baselineskip}\noindent 
  \textbf{\textsf{\small DATA AVAILABILITY}}
  \par\vspace{\baselineskip}
}%
{
}

\begin{document}

\preprint{AIP/123-QED}

\title[]{Collective dynamics in the presence of finite-width pulses}

\author{Afifurrahman}
\email{r01a17@abdn.ac.uk}

\author{Ekkehard Ullner}%
 \email{e.ullner@abdn.ac.uk}

\author{Antonio Politi}
\email{a.politi@abdn.ac.uk}

\affiliation{%
Institute for Pure and Applied Mathematics and Department of Physics (SUPA), Old Aberdeen, Aberdeen, AB24 3UE, United Kingdom 
}%

\date{\today}

\begin{abstract}

The idealisation of neuronal pulses as $\delta$-spikes is a convenient approach in neuroscience but can sometimes lead to erroneous conclusions. We investigate the effect of a finite pulse-width on the dynamics of balanced neuronal networks.
In particular, we study two populations of identical excitatory and inhibitory neurons in a random network of phase
oscillators coupled through exponential pulses with different widths. We consider three coupling functions,
inspired by leaky integrate-and-fire neurons with delay and type-I phase-response curves. By exploring the role of the
pulse-widths for different coupling strengths we find a robust collective irregular dynamics, which collapses onto a
fully synchronous regime if the inhibitory pulses are sufficiently wider than the excitatory ones. The transition
to synchrony is accompanied by hysteretic phenomena (i.e. the co-existence of collective irregular and synchronous
dynamics). Our numerical results are supported by a detailed scaling and stability analysis of the fully synchronous solution. A conjectured first-order phase transition emerging for $\delta$-spikes is smoothed out for finite-width
pulses.
\end{abstract}

\maketitle

\begin{quotation}
Neuronal networks with a nearly balanced excitatory/inhibitory activity evoke significant interest in neuroscience due to the resulting emergence of strong fluctuations akin to those observed in the resting state of the
mammalian brain. While most studies are limited to a
$\delta$-like pulse setup, much less is known about the collective behavior in the presence of finite pulse-widths. In
this paper, we investigate exponential pulses, with the goal of testing the
robustness of previously identified regimes such as the
spontaneous emergence of collective irregular dynamics
(CID), an instance of partial synchrony with a non-periodic macroscopic dynamics. Moreover, the finite-width assumption paves the
way to the investigation of a new ingredient, present in
real neuronal networks: the asymmetry between excitatory and inhibitory pulses. Our numerical studies confirm
the emergence of CID also in the presence of finite pulse-width, although with a couple of warnings: (i) the amplitude of the collective fluctuations decreases significantly
when the pulse-width is comparable to the interspike interval; (ii) CID collapses onto a fully synchronous regime
when the inhibitory pulses are sufficient longer than the
excitatory ones. Both restrictions are compatible
with the recorded behavior of real neurons.
Additionally, we find that a seemingly first-order phase
transition to a (quasi)-synchronous regime disappears in the presence of a finite width, 
confirming the peculiarity of the $\delta$-spikes. A transition
to synchrony is instead observed upon increasing the ratio
between the width of inhibitory and excitatory pulses: this
transition is accompanied by a hysteretic region, which
shrinks upon increasing the network size. Interestingly,
for a connectivity comparable to that of the mammalian
brain, such a finite-size effect is still sizable. Our numerical studies might help to understand abnormal synchronisation in neurological disorders.
\end{quotation}

\section{Introduction}
Irregular firing activity is a robust phenomenon observed in certain areas of mammalian brain, such as hippocampus or 
cortical neurons~\cite{Jarosiewicz2002,Shinomoto2009}. It plays a key role for the brain functioning 
in the visual and prefrontal cortex. 
This behavior emerges from the combined action of many interacting units~\cite{Truccolo2009,Gerstner2014}.

This paper focuses on a regime called {\it collective irregular dynamics} (CID), which arises in networks of
oscillators (neurons).
Mathematically, CID is a non-trivial form of  partial synchrony. Like partial synchrony, it means that the
order parameter $\chi$ used to identify synchronization (see Sec.~II for a precise definition) 
is strictly larger than 0 and smaller than 1. Moreover, it implies a stochastic like behavior of macroscopic observables such as the average membrane potential.

There are (at least) two mechanisms leading to CID: (i) the intrinsic infinite dimensionality of the 
nonlinear equations describing whole populations of oscillators; (ii)
an imperfect balance between excitatory and inhibitory activity.

Within the former framework, no truly complex collective dynamics can arise 
in mean-field models of identical oscillators of Kuramoto type. In fact, the Ott-Antonsen Ansatz~\cite{OA2008} 
implies a strong dimension reduction of the original equations.
Nevertheless, in this and similar contexts, CID can arise either in the presence of a delayed feedback~\cite{Pazo2016}, or when two interacting populations are considered~\cite{Olmi2010}. 
Alternatively, it is sufficient to consider either ensembles of heterogeneous oscillators: e.g.,
leaky integrate-and-fire (LIF) neurons~\cite{Luccioli2010}, and
pulse-coupled phase oscillators~\cite{ullner_politi2016}
(notice that in these cases, Ott-Antonsen Ansatz does not apply).

Within the latter framework, an irregular activity was first observed and described in networks of binary units,
as a consequence of a (statistical) balance between excitation and inhibition~\cite{Sompolinsky1996}. 
This {\it balanced} regime~\cite{Vreeswijk1998} can be seen as an asynchronous state accompanied by statistical fluctuations.
In fact, this interpretation led Brunel~\cite{Brunel2000} to develop a powerful analytical method based on a 
self-consistent Fokker-Planck equation to describe an ensemble of LIF neurons.
In the typical (sparse) setups considered in the literature, the fluctuations of the single neuron activity vanish
when averaged over the whole population, testifying to their statistical independence; in terms of order parameter,
$\chi=0$.

However, it has been recently shown that a truly CID can be observed in the presence of massive coupling 
(finite connectivity-density) under the condition of small unbalance~\cite{ullner2018,Politi2018}. 
In this paper we test the robustness of these results, obtained while dealing with $\delta$-pulses,
by studying more realistic finite-width pulses.
In fact, real pulses have a small but finite width~\cite{Canavier2013}. Moreover, it has been shown
that the stability of some synchronous regimes of LIF neurons may qualitatively change, when arbitrarily short pulses are
considered (in the thermodynamic limit)~\cite{Politi2007}. 

A preliminary study has been already published in Ref.~[\citen{Zhou2017}], where the authors have 
not performed any finite-size scaling analysis and, more important, no any test of the presence of CID has been carried out. 
Here we study a system composed of two
populations of (identical) excitatory and inhibitory neurons, which interact via exponential pulses of different width, as it
happens in real neurons~\cite{Gerstner-Kistler-02}.

Handling pulses with a finite width requires two additional variables per single neuron, in order to describe
the evolution of the incoming excitatory and inhibitory fields. 
The corresponding mathematical setup has been recently studied in Ref.~[\citen{af_eu_ap2020}] with the goal of determining the stability of
the fully synchronous state in a sparse network. The 
presence of two different pulse-widths leads to non-intuitive stability properties, because the different time dependence of the two pulses
may change the excitatory/inhibitory character of the overall field perceived by each single neuron.
Here, we basically follow the same setup introduced in Ref.~[\citen{af_eu_ap2020}] with the main difference of a massively coupled network,
to be able to perform a comparative analysis of CID.

The randomness of the network accompanied by the presence of three variables per neuron, makes an analytical treatment quite
challenging. For this reason
we limit ourselves to a numerical analysis. However, we accompany our studies with a careful finite-size scaling to extrapolate
the behavior of more realistic (larger) networks. 
Our first result is that CID is observed also in the presence of finite pulse-width, although 
we also find a transition to full synchrony when the inhibitory pulses are sufficiently longer than excitatory ones.
The transition is first-order (discontinuous) and is accompanied by hysteresis: there exists a finite range
of pulse-widths where CID and synchrony coexist.

The finite-size analysis suggests that in the thermodynamic limit CID is not stable 
when the pulses emitted by inhibitory neurons are strictly
longer than those emitted by the excitatory ones. However, the convergence is rather slow and we cannot exclude that the asymmetry plays an
important role in real neuronal networks of finite size.

More precisely in section II, 
we define the model, including the phase response curves (PRCs) used in our numerical simulation. 
In the same section we also introduce the tools and indicators later used to characterize the dynamical regimes, notably an order
parameter to quantify the degree of synchronization~\cite{GOLOMB2001}. 
In section III, we present some results obtained for strictly $\delta$ pulses to test robustness of CID in our context of coupled phase oscillators.
In Sec. IV we discuss the symmetric cases of identical finite pulse-widths. Sec. V is devoted to a thorough analysis of CID by varying the pulse-widths. Sec. VI contains a discussion of the transition
region,  intermediate between standard CID and full synchrony. 
In the same section, the robustness of the transition region is analysed, by considering different PRCs.
Finally, section VII is devoted to the conclusions and a brief survey of the open problems.

\section{\label{sec:model}Model}
Our object of study is a network of $N$ phase oscillators (also referred to as neurons), the first $N_e=bN$ 
being excitatory, the last $N_i=(1-b)N$ inhibitory (obviously, $N_e+N_i = N$).
Each neuron is characterized by the phase-like variable $\Phi^j\le 1$ (formally equivalent to a membrane 
potential), while the (directed) synaptic connections are represented  by the connectivity matrix $\mathbf{G}$ with entries
\begin{align*}
& G_{j,k}=
    \begin{cases}
    1, & \text{if $k \to j$ active} \\
    0, & \text{otherwise } 
    \end{cases}
\end{align*}
where $\sum_{k = 1}^{N_e} G_{j,k} = K_e$ and
$\sum_{k = N_e+1}^{N} G_{j,k} =K_i$, meaning that each neuron $j$ is characterized by the same number of incoming
excitatory and inhibitory connections, as customary assumed in the literature \cite{ostojic2014} 
($K=K_e+K_i$ represents the connectivity altogether).  Here, we assume that $K$ is proportional to $N$, that is $K=cN$, i.e. we refer to
massive connectivity. Further, the network structure is without autapse, i.e. $G_{j,j} = 0$.

The evolution of the phase of both excitatory and inhibitory neurons  is ruled by the same equation, 
\begin{equation}
   \dot{\Phi}^j  =   1 + \frac{\mu}{\sqrt{K}} \, \Gamma\left(\Phi^j\right)\left(E^j - I^j\right)\label{eq:mod1} \, ,
\end{equation}
where $E^j(I^j)$ the excitatory (inhibitory) field perceived by the $j$th neuron, while
$\Gamma(\Phi)$ represents the phase-response curve (PRC) assumed equal for all neurons; finally,
$\mu$ is the coupling strength. 
Whenever $\Phi^k$ reaches the threshold $\Phi_{th}=1$, it is reset to the value $\Phi_r = 0$ and 
enters a refractory period $t_{r}$ during which it stands still and is insensitive to the action of both fields.
The fields $E^j$ and $I^j$ are the linear superposition of exponential spikes emitted by the upstream neurons.
Mathematically,
\begin{eqnarray}
\dot{E}^j    &=& -\alpha\left( E^j - \sum_{n}  G_{j,k} P_{k} \delta(t-t^k_n) \right)  \label{eq:mod2} \\
\dot{I}^j    &=& -\beta \left( I^j - g\sum_{n} G_{j,k} (1-P_{k})\delta(t-t^k_n) \right) \, ,\nonumber
\end{eqnarray}
where $\alpha$ ($\beta$) denotes the inverse pulse-width of the excitatory (inhibitory) spikes and $t^k_n$ is the emission time of the $n$th spike emitted by the $k$th neuron. 
The coefficient $g$ accounts for the relative strength of inhibition compared to excitation. 
If the $k$th neuron is excitatory, $P_{k}=1$, otherwise $P_k =0$.

In the limit of $\alpha(\beta) \rightarrow \infty$ ($\delta$-spikes) both fields can be expressed as simple sums 
\begin{eqnarray}
E^j    &=&  \sum_{n}  G_{j,k} P_{,k} \delta(t-t^k_n)   \label{eq:mod3} \\
I^j    &=&  g\sum_{n} G_{j,k} (1-P_{k})\delta(t-t^k_n) \, .\nonumber
\end{eqnarray}

Let us now introduce the PRCs used later in our numerical simulations. We consider three different shapes:
\begin{itemize}
\item PRC\textsubscript{1}
\begin{equation}\label{eq:prci}
  \Gamma(\Phi^j) =
  \begin{cases}
  \left(\Phi^j - \Phi_L \right) & \text{if $\Phi_L<\Phi^j <\Phi_U$} \\
  0  & \text{otherwise}
  \end{cases}
\end{equation}

\item PRC\textsubscript{2}
\begin{equation}\label{eq:prcii}
  \Gamma(\Phi^j) =
  \begin{cases}
  \frac{\Phi^j-\Phi_L}{0.5-\Phi_L} & \text{if $\Phi_L<\Phi^j < 0.5$} \\
  1 - \left(\frac{\Phi^j - 0.5}{\Phi_U - 0.5} \right) & \text{if $0.5<\Phi^j < \Phi_U$} \\
  0  & \text{otherwise}
  \end{cases}
\end{equation}

\item PRC\textsubscript{3}
\begin{equation}\label{eq:prciii}
\Gamma(\Phi^j)=\sin^2\left(\pi \Phi^j \right)
\end{equation}
\end{itemize}

\begin{figure}
\centering
\includegraphics[width=.47\textwidth]{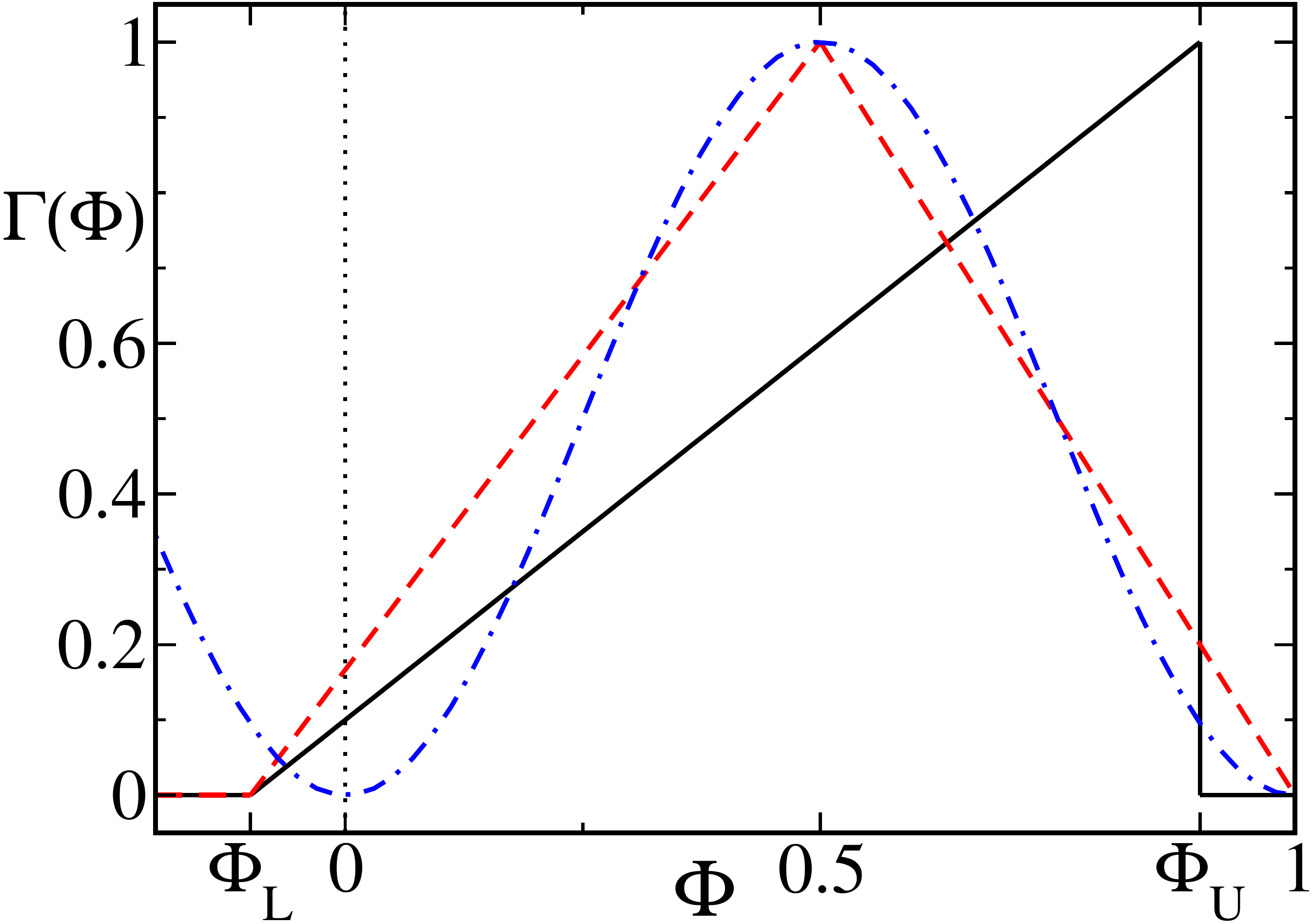}
\caption{Example of the phase response curves (PRCs): PRC\textsubscript{1} with $\Phi_L=-0.1$ and $\Phi_U=0.9$ (black line), PRC\textsubscript{2} (red dashed line), PRC\textsubscript{3} (blue dashed and dot line). The vertical dot line refers to the reset membrane potential ($\Phi_{r} = 0$).}
\label{fig:fig1}
\end{figure}

The various curves are plotted in Fig.~\ref{fig:fig1}.
PRC\textsubscript{1} (see the black curve, which corresponds to $\Phi_L = -0.1$ and $\Phi_U = 0.9$) has been introduced
in Ref.~[\citen{af_eu_ap2020}] to study the stability of the synchronous regime; its shape has been proposed to
mimic a network of leaky integrate-and-fire neurons in the presence of delay (see also~Ref.~[\citen{ullner_politi2016}]).

The two other PRCs have been selected so as to explore the effect of a progressive regularization of the neuronal response. 
In particular, we consider the smooth PRC\textsubscript{3} (see Eq.~(\ref{eq:prciii})), as a prototype of 
type I PRC~\cite{Canavier2006,Izhikevich2008}.

The network dynamics is simulated by implementing the Euler algorithm with a time step $\delta_t = 10^{-3}$. However, in some cases,
smaller steps have been considered to avoid spurious synchronization.
We typically initialize the phases uniformly in the unit interval, while the fields are initially set equal to zero.

In all cases, we have set $b=0.8$, $c=0.1$ and $g=4 + \sqrt{1000/K}$ (following the existing literature~\cite{ullner2018}). 
The last condition ensures that the balanced regime is maintained for $K$, $N \to \infty$.
Moreover, we have systematically explored the role of $\alpha$ and $\beta$, as the pulse-width is the focal point of this
paper. Additionally, the coupling strength $\mu$ has been varied, as well as the network-size $N$, to test for
the amplitude of finite-size effects.

The following statistical quantities are used to characterize the emerging dynamical states.

\begin{enumerate}

\item \textit{The mean firing rate} is a widely used indicator to quantify the neural activity.
It is defined as 
\begin{equation}\label{5}
    \nu = \lim_{t \rightarrow\infty}\frac{1}{t N}  \sum_{j=1}^N \mathcal{N}_j(t)
\end{equation}
where $\mathcal{N}_j(t)$ denotes the number of spikes emitted by the neuron $j$ over a time $t$.

\item \textit{The coefficient of variations} $C_v$ is a microscopic measure of irregularity 
of the dynamics based on the fluctuations of the interspike intervals (ISIs). 
The average $C_v$ is defined as
\begin{equation}\label{6}
    \langle{C}_v\rangle= \frac{1}{N} \sum_{j=1}^N \frac{\sigma_j}{\tau_j},
\end{equation}
where $\sigma_j$ is the standard deviation of the single-oscillator's ISI, and $\tau_j$ is the corresponding mean ISI. If $\langle{C}_v\rangle>1$, then the neurons show a bursting activity, while $\langle{C}_v\rangle<1$ means that the spike train is relatively regular.

\item \textit{The order parameter}, $\chi$, is typically used to quantify the degree of synchronisation of a population
of neurons~\cite{Golomb2007}. It is defined as
\begin{equation}\label{3}
    \chi^2 \equiv \frac{\overline{{\langle \Phi \rangle}^2} - \overline{{\langle \Phi \rangle}}^2}{\langle \overline{\Phi^2} - \overline{\Phi}^2 \rangle},
\end{equation}
where $\langle \cdot \rangle$ represents an ensemble average, while the over-bar is a time average.
The numerator is the variance of the ensemble average $\langle \Phi \rangle$, while the denominator is the ensemble mean of the
single-neuron's variances.
When all neurons behave in exactly the same way (perfect synchronization), then $\chi=1$. 
If instead, they are independent, then $\chi \approx 1/\sqrt{N}$. 
Regimes characterized by intermediate finite values $0<\chi<1$ are referred to as instances of partial synchronization. 
However, $\chi >0$ does not necessarily imply that the collective dynamics is irregular: it is, e.g., compatible with
a periodic evolution. In fact, here we report several power spectra to testify the stochastic-like dynamics of 
macroscopic (average) observables.
\end{enumerate}

\section{Delta Pulse}
\begin{figure}
\centering
\includegraphics[width=.47\textwidth]{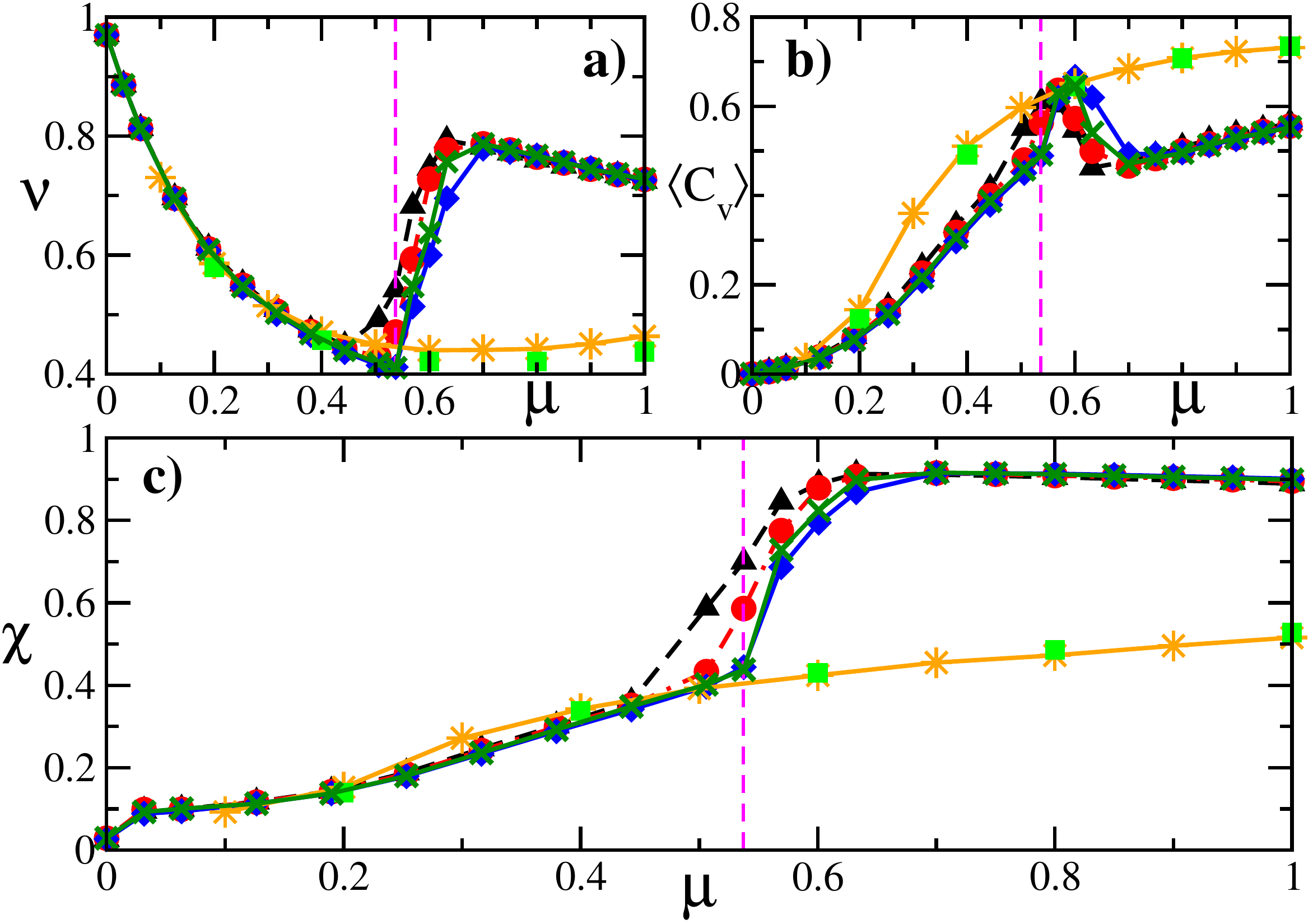}
\caption{Characterization of the global network dynamics with interactions through $\delta$-pulses. 
Mean firing rate $\nu$, mean coefficient of variations $\langle{C}_v\rangle$, and order parameter $\chi$
are plotted vs. the coupling strength $\mu$ in panels (a), (b) and (c), respectively. 
Black triangles, red circles, green crosses, and blue diamonds correspond to $N=10000$, 20000, 40000, and 80000, respectively,
all obtained with PRC\textsubscript{1}.
Orange stars and green squares correspond to $N=10000$ and 40000 obtained with PRC\textsubscript{3}.
The vertical dashed line represents the critical coupling $\mu_c = 0.537$.}
\label{fig:fig2}
\end{figure}

Most spiking network models deal with $\delta$-spikes, including those giving rise to CID\cite{ullner2018,Politi2018}. 
This paper is focused on the more realistic exponential spikes, but before proceeding in that direction 
we wish to briefly discuss the case of zero pulse-width. 
This is useful to gauge the different PRCs used in this paper. Since $\delta$ pulses correspond to the limiting case $\alpha, \beta \to \infty$, they can be treated by invoking Eq.~(\ref{eq:mod3}). 
Figure~\ref{fig:fig2} shows the various indicators introduced in Section II, to characterize the collective dynamics. 
As in previous papers,\cite{ullner2018,Politi2018} we explore the parameter space, by varying
the coupling strength ${\mu}$ and the system size $N$.

In panel (c) we can appreciate that CID emerges already for very small coupling strength; it is accompanied by an
increasing average coefficient of variations $\langle{C}_v\rangle$, due to the coupling which induces increasing deviations from
purely periodic behavior. In parallel, the mean firing rate $\nu$ decreases as a result of the prevalent inhibitory
character of the network.
This weak-coupling emergence of CID is comparable to what observed in balanced LIF models with $\delta$ spikes\cite{ullner2018}. 

Above $\mu \approx 0.537 \equiv \mu_{c}$ (see the vertical dashed lines), a transition occurs towards a
highly synchronous regime ($\chi$ is slightly smaller than 1), accompanied by a larger firing rate.
The corresponding firing activity is mildly irregular: $\langle{C}_v\rangle$ is smaller than in
Poisson processes (when $\langle{C}_v\rangle=1$). A quick analysis suggests that this self-sustained regime emerges from
the vanishing width of the pulses combined with the PRC shape, which is strictly equal to zero in a finite phase range
below the threshold $\Phi_{th}=1$. In fact, similar studies performed with PRC\textsubscript{3} do not reveal any evidence of a phase
transition (see orange stars and green squares in Fig.~\ref{fig:fig2}) indicating that such behavior is nothing else but 
a peculiarity of PRC\textsubscript{1} with $\delta$-pulses. We have not further explored this regime.
It is nevertheless worth noting that the sudden increase of the firing rate observed when passing to the strong coupling
regime is reminiscent of the growth observed in LIF neurons~\cite{ostojic2014}, although in such a case, the increase 
is accompanied by a significantly bursty behavior~\cite{eu_ap_at20}.

More important is the outcome of the finite-size scaling analysis, performed to investigate the robustness of the observed scenario.
In Fig.~\ref{fig:fig2} one can see that the various indicators under stimulation of PRC\textsubscript{1} are size-independent deeply within the two dynamical phases, 
while appreciable deviations are observed in the transition region. This is customary when dealing with phase-transitions.
It is not easy to conclude whether the transition is either first or second order: the $\langle{C}_v\rangle$ is reminiscent of
the divergence of susceptibility seen in continuous transitions, but this is an issue that would require additional work
to be assessed.

\section{Identical finite-width pulses}
In this section, we start our analysis of finite pulses, by assuming the same width for inhibitory and excitatory neurons,
i.e. $\alpha^{-1}=\beta^{-1}$. The asymmetric case is discussed in the next section.
All other system parameters are kept the same as in the previous section (including the PRC shape).


\begin{figure*}
\centering
\includegraphics[width=0.8\textwidth]{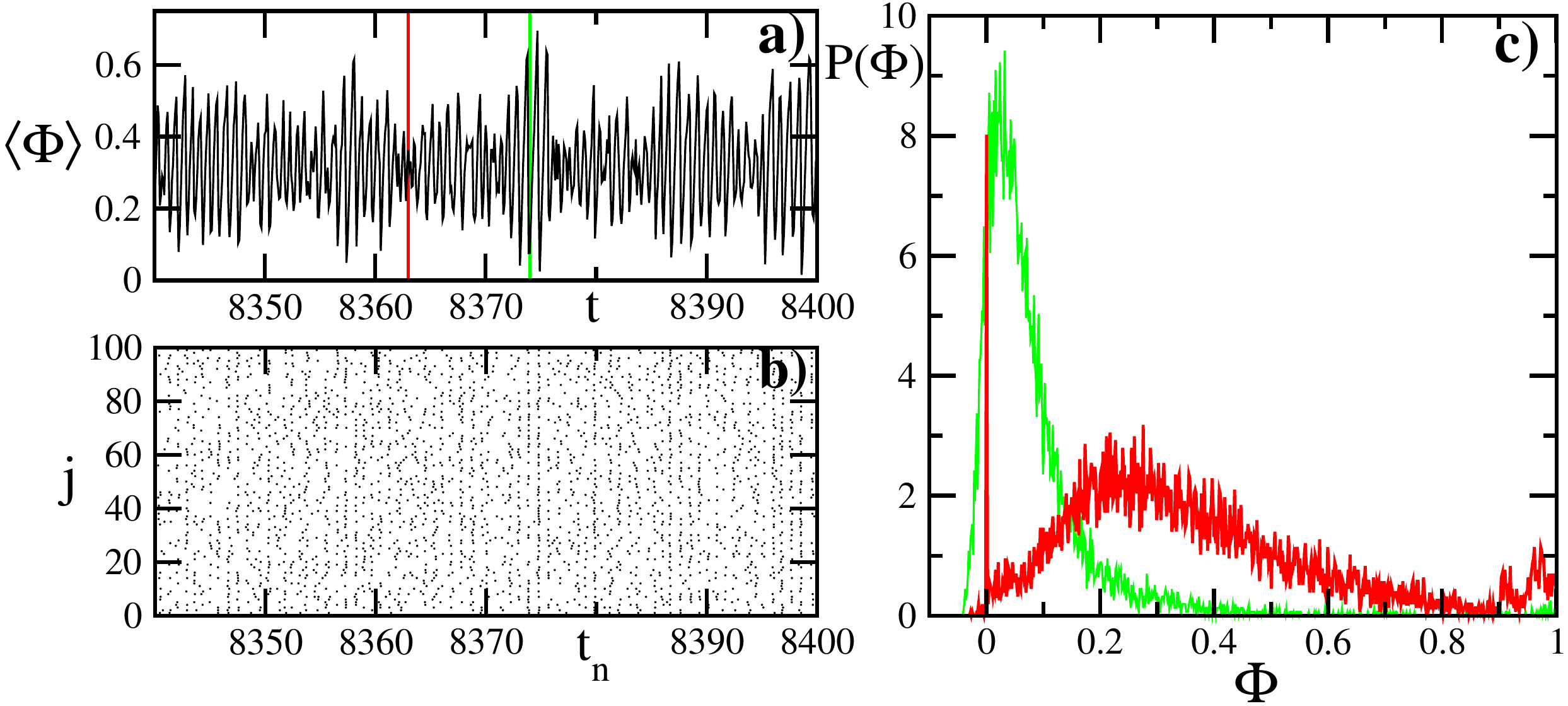}
\caption{CID properties for PRC\textsubscript{1}, $\mu=0.95$, $\alpha = \beta=100$, and $N=10000$. 
Panel a): time series of the mean field $\langle \Phi \rangle$. 
Panel b): raster plot of spiking times $t_n$ for 100 oscillators out of $N=10000$. 
Panel c): instantaneous probability distribution of the phases P$(\Phi)$ at two different time points $t=8363$ (red), and $t=8374$ (green). The probability distributions are normalized such that the area underneath is 1.}
\label{fig:fig3}
\end{figure*}

Before discussing the macroscopic measures, we turn our attention to typical CID features.
The average phase $\langle \Phi \rangle (t) =  \frac{1}{N}  \sum_j \Phi_j (t)$ (see Fig.~\ref{fig:fig3}(a)) exhibits stochastic-like oscillations, which represent a first
evidence of a non-trivial collective dynamics. 
The raster plot presented in panel Fig.~\ref{fig:fig3}(b) contains the firing times $t_n$ of a subset of 100 neurons:
there, one can easily spot the time intervals characterized by a more coordinated action
(see, for instance, around the vertical green line at time $8374$ in Fig.~\ref{fig:fig3}(a)).
A more quantitative representation is presented in Fig.~\ref{fig:fig3}(c), where 
the instantaneous phase distribution $P(\Phi)$ is plotted at two different times in correspondence of  qualitatively
different regimes of the phase dynamics (see the vertical lines in panel (a)).
The peak at $\Phi = 0$ is due to the finite fraction of neurons standing still in the refractory period. 
A small amount of negative phases are also seen: they are due to prevalence of inhibition over excitation
at the end of refractoriness. 
Moreover, the instantaneous phase distribution $P(\Phi)$ presented in Fig.~\ref{fig:fig3}(c), show
that, at variance with the classical asynchronous regime, the shape of the probability density changes with time.
The narrowest distribution (green curve) corresponds to the region where strong regular oscillations 
of $\langle \Phi \rangle$ are visible in panel (a): within this time interval  a ``cloud" of neurons homogeneously
oscillates from reset to threshold and back.


\begin{figure}
\centering
\includegraphics[width=.48\textwidth]{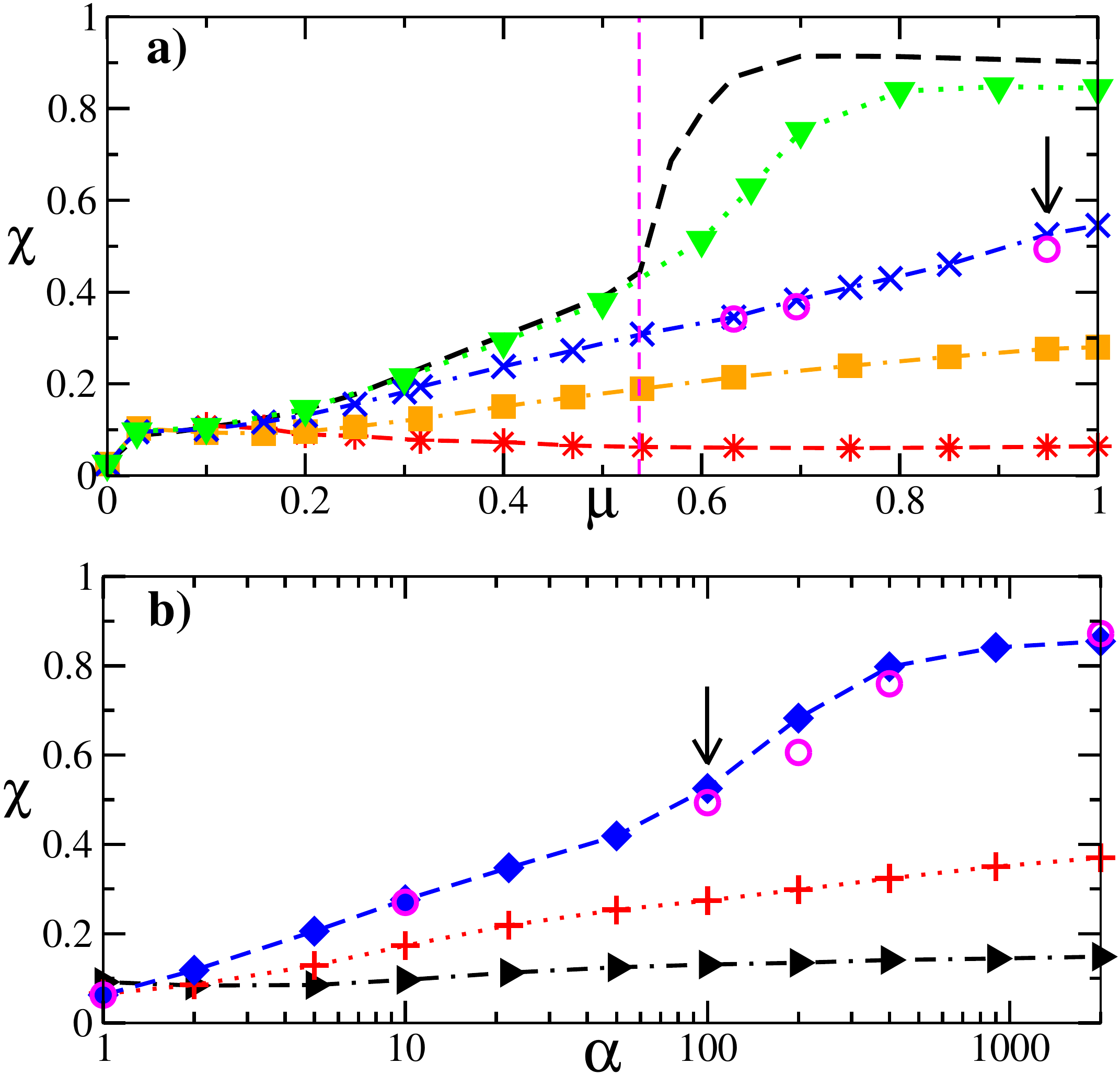}
\caption{Global network dynamics in the presence of identical finite pulse-width and PRC\textsubscript{1}. Panel a): order parameter $\chi$ vs. $\mu$ for $N=10000$ and $\alpha=1000$ (green triangles), $\alpha=100$ (blue crosses), $\alpha=10$ (orange squares), and $\alpha=1$ (red stars). The black dashed curve corresponds to the asymptotic results obtained for $\delta$ pulses (see Fig.~\ref{fig:fig2} (c), $N=80000$) with the critical value $\mu_c$ derived therein. Panel b): order parameter $\chi$ vs. $\alpha$ for $N=10000$ and $\mu=0.2$ (black triangles), $\mu=0.47$ (red pluses), and $\mu=0.95$ (blue diamonds). In both panels the magenta circles show the results for $N=40000$ to compare with the blue curves, respectively. The arrows highlight the parameter set for which we show in Fig.~\ref{fig:fig3} typical CID time series.}
\label{fig:fig4}
\end{figure}

The resulting order parameter is reported in Fig.~\ref{fig:fig4}. In panel (a) we plot $\chi$ 
as a function of $\mu$ for different widths: from broad pulses (red stars correspond to $\alpha =1$, a width comparable to
the ISI), down to very short ones (green triangles correspond to $\alpha=1000$).
The general message is that partial synchrony is preserved.
Nevertheless, it is also evident that increasing the width progressively decreases the amplitude of the order parameter.
The main qualitative difference is the smoothening of the transition observed for $\delta$-pulses (in correspondence of the
vertical dashed line at $\mu_c$). The singular behavior of $\delta$-spikes is confirmed by the relatively large deviations
appearing already for $\alpha = 1000$. 

A more direct illustration of the role of $\alpha$ is presented in Fig.~\ref{fig:fig4}(b), where we plot $\chi$ versus $\alpha$ for
different coupling strengths: $\mu = 0.2$ (black triangles), 0.47 (red crosses), and 0.95 (blue diamonds).
An overall increasing trend upon shortening the pulse-width is visible for all coupling strengths, although the rate 
is relatively modest for weak coupling, becoming more substantial in the strong-coupling limit.

Finally, we have briefly investigated the presence of finite-size effects, by performing some simulations for $N=40000$
(to be compared with $N=10000$ used in the previous simulations): see magenta circles in both panels.
We can safely conclude that the overall scenario is insensitive to the network size.

\section{Full setup}
In the previous section we have seen that the finite width of the spikes does not kill the spontaneous emergence of CID. Here, we
analyse the role of an additional element: the asymmetry between inhibitory and excitatory pulses. We proceed by exploring
the two-dimensional parameter space spanned by the coupling strength $\mu$ and the asymmetry between pulse widths. 
The latter parameter dependence is explored by setting $\alpha=100$ and letting $\beta$ (the inverse width of
inhibitory pulses) vary.
All other network parameters, including the PRC shape, are assumed to be the same as in the previous section. 

\begin{figure*}
\centering
\includegraphics[width=.83\textwidth]{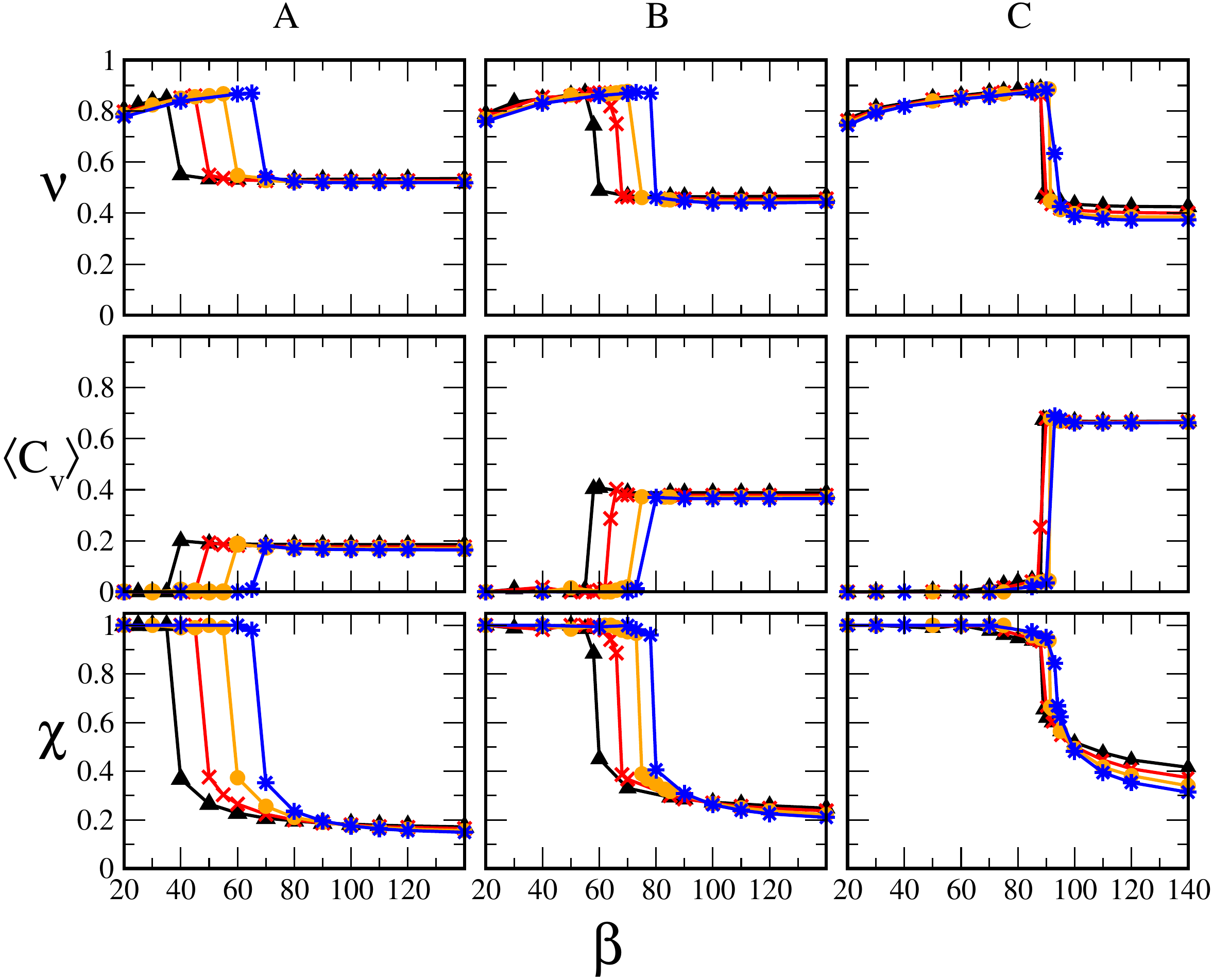}
\caption{Characterization of the global network dynamics for nonidentical finite pulse-width, obtained with $\alpha=100$ and PRC\textsubscript{1}. Each column refers to different coupling strengths: $\mu=0.3$ (A), $\mu=0.47$ (B), and $\mu=0.95$ (C). Rows:  mean firing rate $\nu$, mean coefficient of variations $\langle{C}_v\rangle$, and order parameter $\chi$ versus $\beta$. Colours and symbols define network sizes $N$: 10000 (black triangles), 20000 (red crosses), 40000 (orange circles), and 80000 (blue stars). Each data point is based on a time series generated over 10000 time units and sampled every 1000 steps after the transient has sorted out. }
\label{fig:fig5}
\end{figure*}

The microscopic manifestation of CID in the setup with non-identical pulses is qualitatively the same as for identical pulses shown in Fig.~\ref{fig:fig3}.
The results of a systematic numerical analysis are plotted in Fig.~\ref{fig:fig5}, where we report three indicators: 
the firing rate $\nu$, the mean coefficient of variation $\langle{C}_v\rangle$, and
the order parameter $\chi$, versus $\beta$ for three different coupling strengths (see the different columns), and
four network sizes.

All indicators reveal the existence of two distinct phases: a synchronous regime arising for small $\beta$
values, and CID observed beyond a critical point which depends on the network size: the transition is discontinuous.
All panels reveal a substantial independence of the network size, with the exception of the transition between them
(we further comment on this issue later in this section). 

The first regime is synchronous and periodic, as signalled by $\chi = 1$, and $\langle{C}_v\rangle=0$.
The corresponding firing rate $\nu$ is a bit smaller than 0.97, the rate of uncoupled neurons 
(taking into account refractoriness). 
This is consistent with the expected predominance of inhibition over excitation in this type of setup. 
A closer look shows that in the synchronous regime $\nu$ increases with $\beta$. This makes sense since the smaller $\beta$, 
the longer the time when inhibition prevails thereby decreasing the network spiking activity.     
The  weak dependence of $\nu$ on the coupling strength $\mu$ is a consequence of small effective fields felt by neurons when the 
PRC is small.
Finally, for intermediate $\beta$ values (around 80) and large coupling strengths, $\chi$ is large but 
clearly smaller than 1. This third type of regime will be discussed in the next section.

CID is characterized by a significantly smaller order parameter which, generally tends to increase
with the coupling strength. CID is also characterized by a significantly smaller firing rate. This is due the prevalence of
inhibition which is not diminished by the refractoriness as in the synchronous regime.
Finally, the coefficient of variation is strictly larger than 0, but significantly smaller than 1 (the value
of Poisson processes) revealing a limited irregularity of the microscopic dynamics.
In agreement with our previous observations for $\delta$-spikes, $\langle{C}_v\rangle$ increases with the coupling
strength.

Our finite-size scaling analysis also shows that
the degree of asymmetry (between pulse widths) compatible with CID progressively
reduces upon increasing the number of neurons. 
Although the $N$-dependence varies significantly with the
coupling strength, it is natural to conjecture that, asymptotically, CID survives only for $\beta \ge \alpha$.
This is not too surprising from the point of view of self-sustained balanced states. They are expected to survive
only when inhibition and excitation compensate each other: the presence of different time scales makes it difficult, if
not impossible to ensure a steady balance.

Transition to synchrony upon lowering $\beta$ was already observed in Ref.~[\citen{Zhou2017}] in a numerical study
of LIF neurons, where, however, no finite scaling analysis was performed.
Interestingly, the onset of a synchronous activity when inhibition is slower than excitation is also consistent with
experimental observations~\cite{Wang2010}.

\begin{figure}
\centering
\includegraphics[width=.47\textwidth]{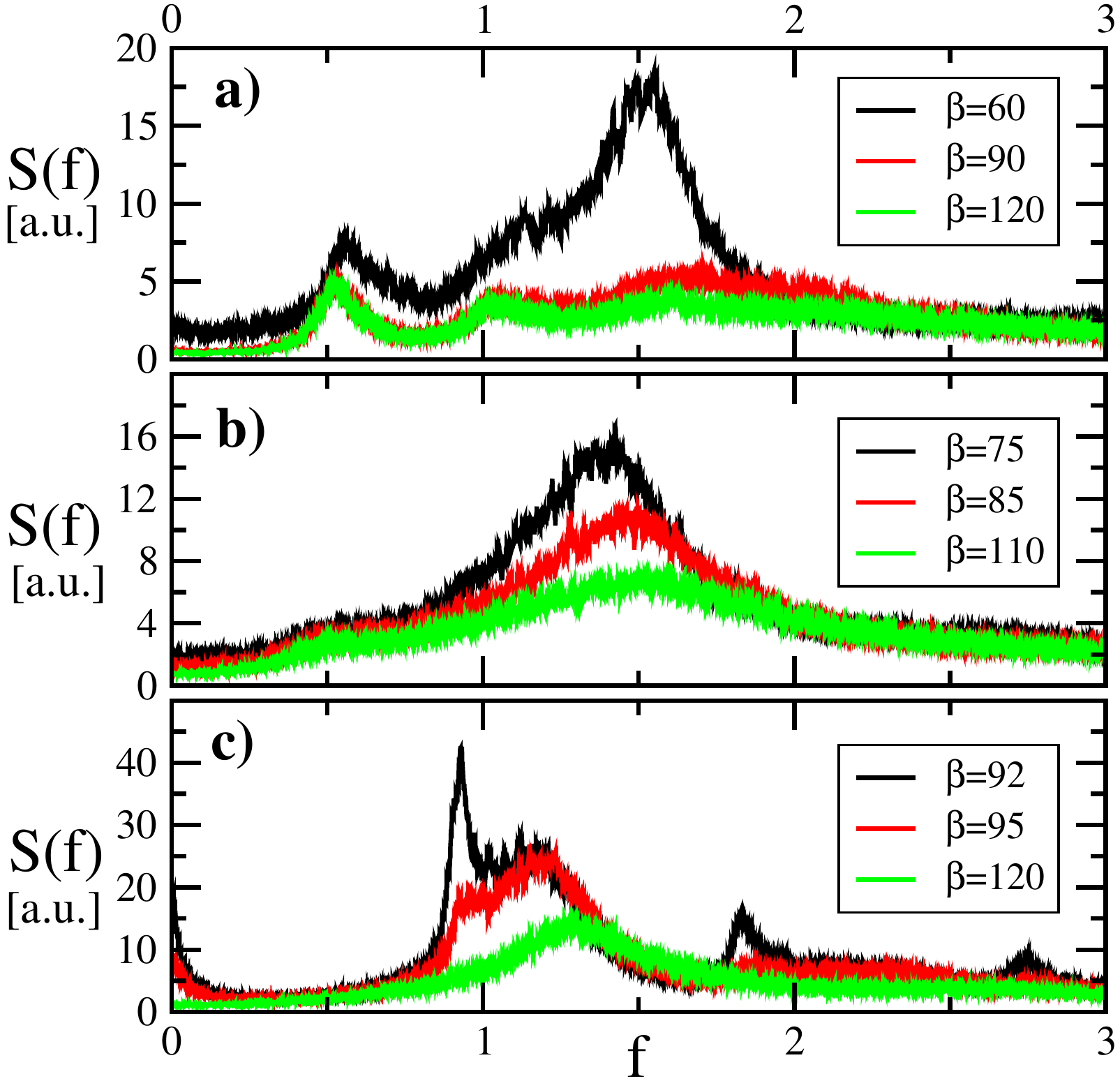}
\caption{Power spectra S(f) of average phase as function of frequency. All presented data refers to PRC\textsubscript{1}, $\alpha=100$ and $N=40000$. Each panel corresponds to different $\mu$: 0.3 (a), 0.47 (b), and 0.95 (c). }
\label{fig:fig6}
\end{figure}

We conclude this section with a more quantitative characterization of the irregularity of the collective dynamics.
In Fig.~\ref{fig:fig6}, we plot the Fourier power spectrum $S(f)$ obtained from $\langle \Phi \rangle (t)$. 
The panels correspond to three different coupling strengths ($\mu = 0.3$, 0.47 and 0.95, from top to bottom). 
For each value of $\mu$, we have sampled three different pulse-widths.

Altogether, one can notice a general increase of the power with $\mu$. This is quite intuitive, as CID is the result 
of mutual interactions. A less obvious phenomenon is the increase of the power observed when the inhibitory pulse-width $\beta^{-1}$
is increased.
This is an early signature of a transition towards full synchronization, observed when $\beta$ is decreased below a
critical value.
Anyway, the most important message conveyed by  Fig.~\ref{fig:fig6} is that all spectra exhibit a broadband structure, 
although most of the power is concentrated around a specific frequency:
$f\approx 1.5$ (panel a),  $f\approx 1.4$ (panel b), and $f\approx 0.93$ (panel c).  
As a result, one can preliminarily conclude that the underlying macroscopic evolution is stochastic-like. A more detailed analysis 
could be performed by computing macroscopic Lyapunov exponents, but this is an utterly difficult task, as it is not 
even clear what a kind of equation one should refer to. 

\begin{figure}
\centering
\includegraphics[width=.47\textwidth]{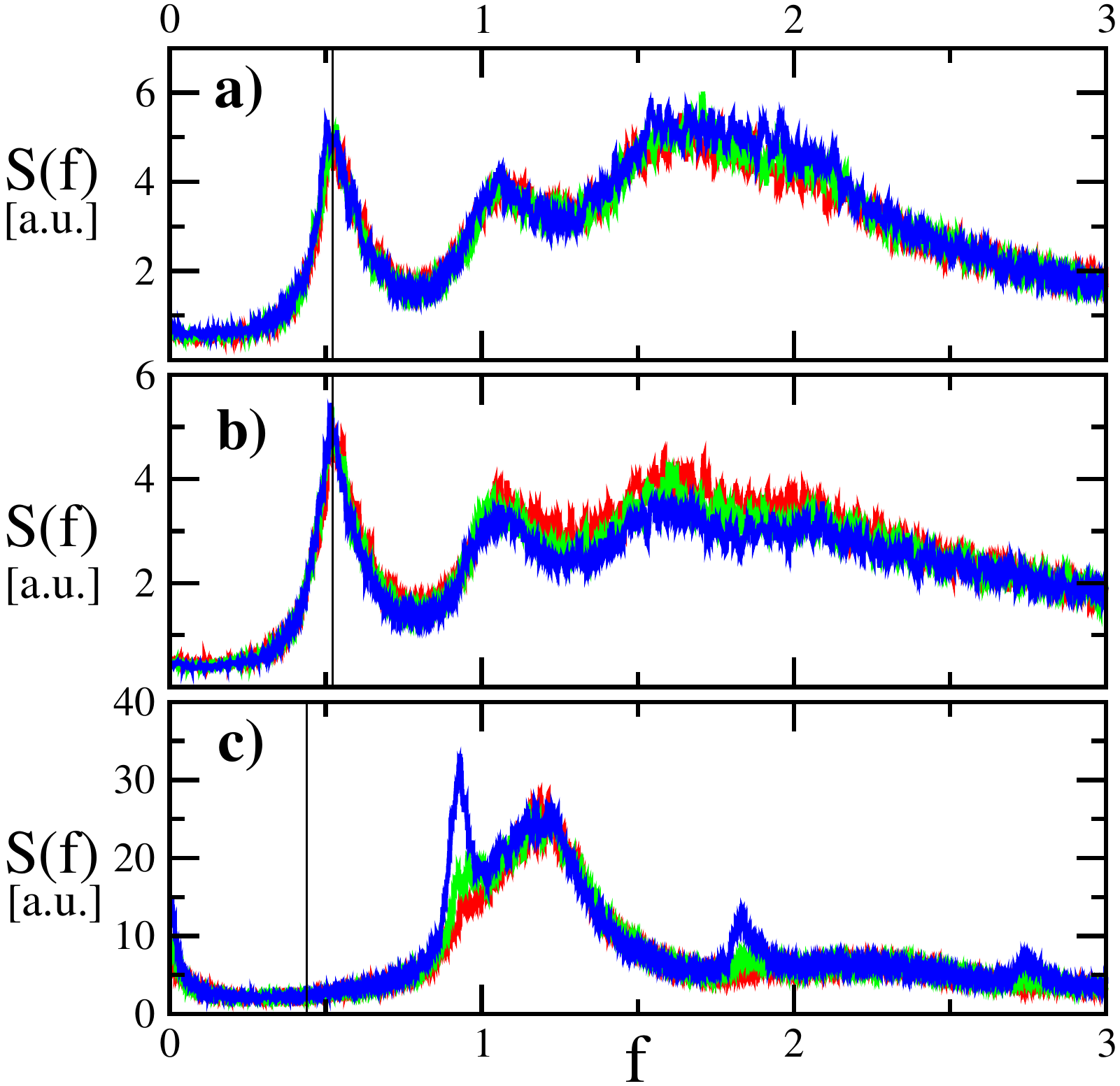}
\caption{Power spectra S(f) of average phase as function of frequency. Panel: a) $\beta=90$, $\mu=0.3$, b) $\beta=120$, $\mu=0.3$ and c) $\beta=95$, $\mu=0.95$. The colour defines network size $N$: 20000 (red), 40000 (green), and 80000 (blue). All presented data refers to PRC\textsubscript{1} and $\alpha=100$. The vertical line is pointing out the mean firing rate $\nu\approx 0.523$ for $\mu=0.3$ and $\nu\approx 0.44$ for $\mu=0.95$ (see Fig.\ref{fig:fig5}).}
\label{fig:fig7}
\end{figure}

Additional evidence of the robustness of CID is given in Fig.~\ref{fig:fig7}, where we investigate
the amplitude of finite-size corrections, by computing the power spectrum $S(f)$ for different network sizes 
for three different parameter sets:
$\mu=0.3$, $\beta=90$ (panel a), $\mu=0.3$, $\beta=120$ (panel b), and $\mu=0.95$, $\beta=95$ (panel c).
In all cases, the spectra are substantially independent of the number of neurons, although in panel (b) we observe a weak 
decrease in the band $f\in [1, 2.5]$, 
while a new set of peaks is born in panel (c). Since the connectivity $K$ of the largest networks herein considered ($N=80000$) 
is comparable to that of the mammalian brain ($K=8000$ vs 10000)\citep{Gerstner2014}, we can at least conjecture that this phenomenon may have
some relevance in realistic conditions.

Finally, the low frequency peak clearly visible for small $\mu$ coincides with the mean firing rate (see Fig.~\ref{fig:fig5}(a)),
while the connection with the microscopic firing rate is lost in panel (c).

\section{Transition region}
In Fig.~\ref{fig:fig5} we have seen a clear evidence of a first-order phase transition, when either the pulse-width or the
coupling strength is varied. 
So far, each simulation has been performed by selecting afresh a network structure.
The stability of our results indicates that the transition does not suffer appreciable sample-to-sample fluctuations.

\begin{figure}
\centering
\includegraphics[width=.48\textwidth]{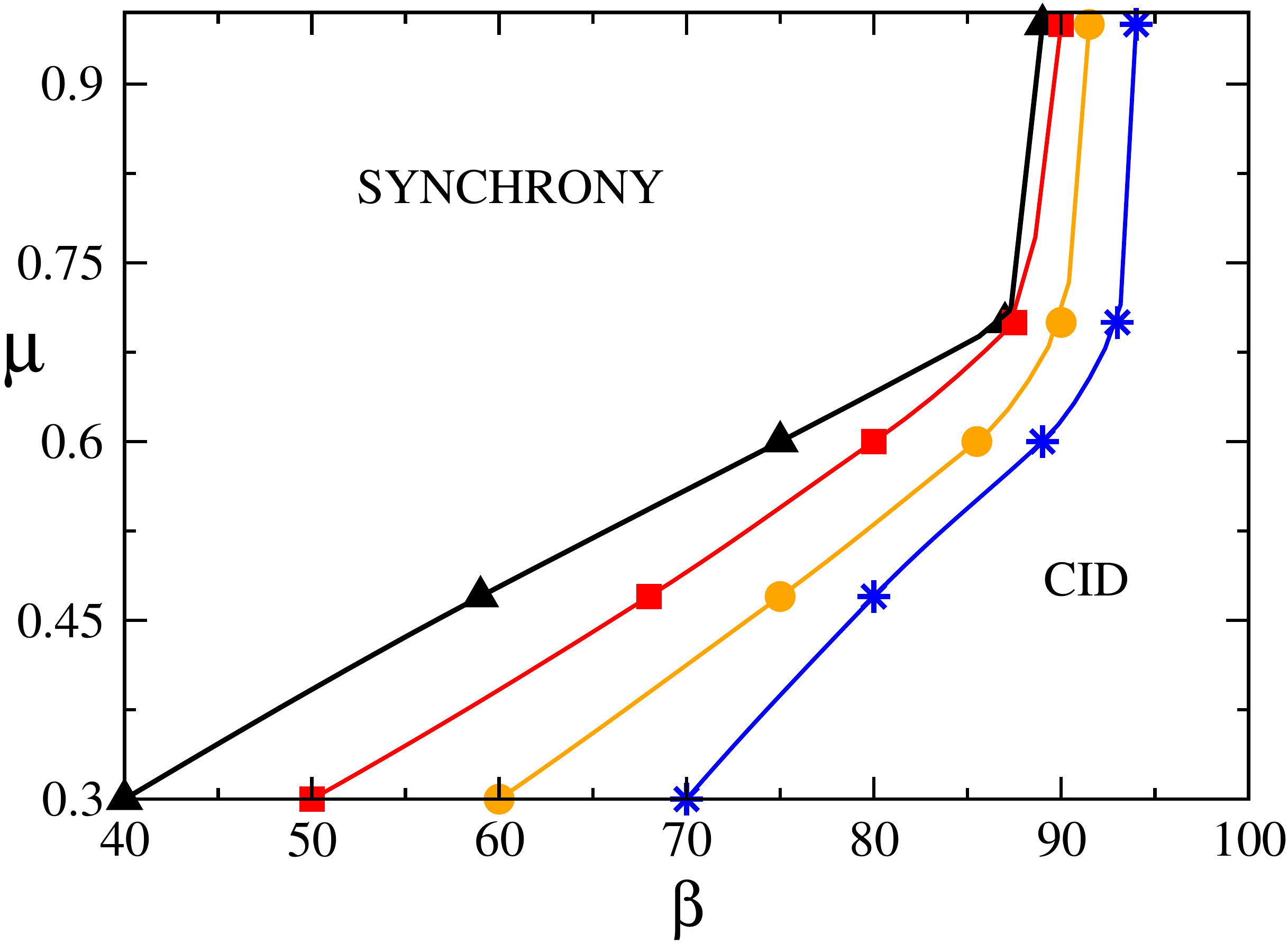}
\caption{Phase diagram obtained with $\alpha=100$ and PRC\textsubscript{1}, for various $N$: 10000 (black triangles), 20000 (red squares), 40000 (orange circles), and 80000 (blue stars).}
\label{fig:fig8}
\end{figure}

The main outcome of our numerical simulations is summarized in Fig.~\ref{fig:fig8}; the various
lines identify the transition between the two regimes, for different network sizes. 
The critical points have been determined by progressively decreasing $\beta$ (see Fig.~\ref{fig:fig5}) and
thereby determining the minimum $\beta$-value where CID is stable. 
Upon increasing $N$, the synchronization region grows and the transition moves towards $\beta=\alpha$.

\begin{figure*}
\centering
\includegraphics[width=.73\textwidth]{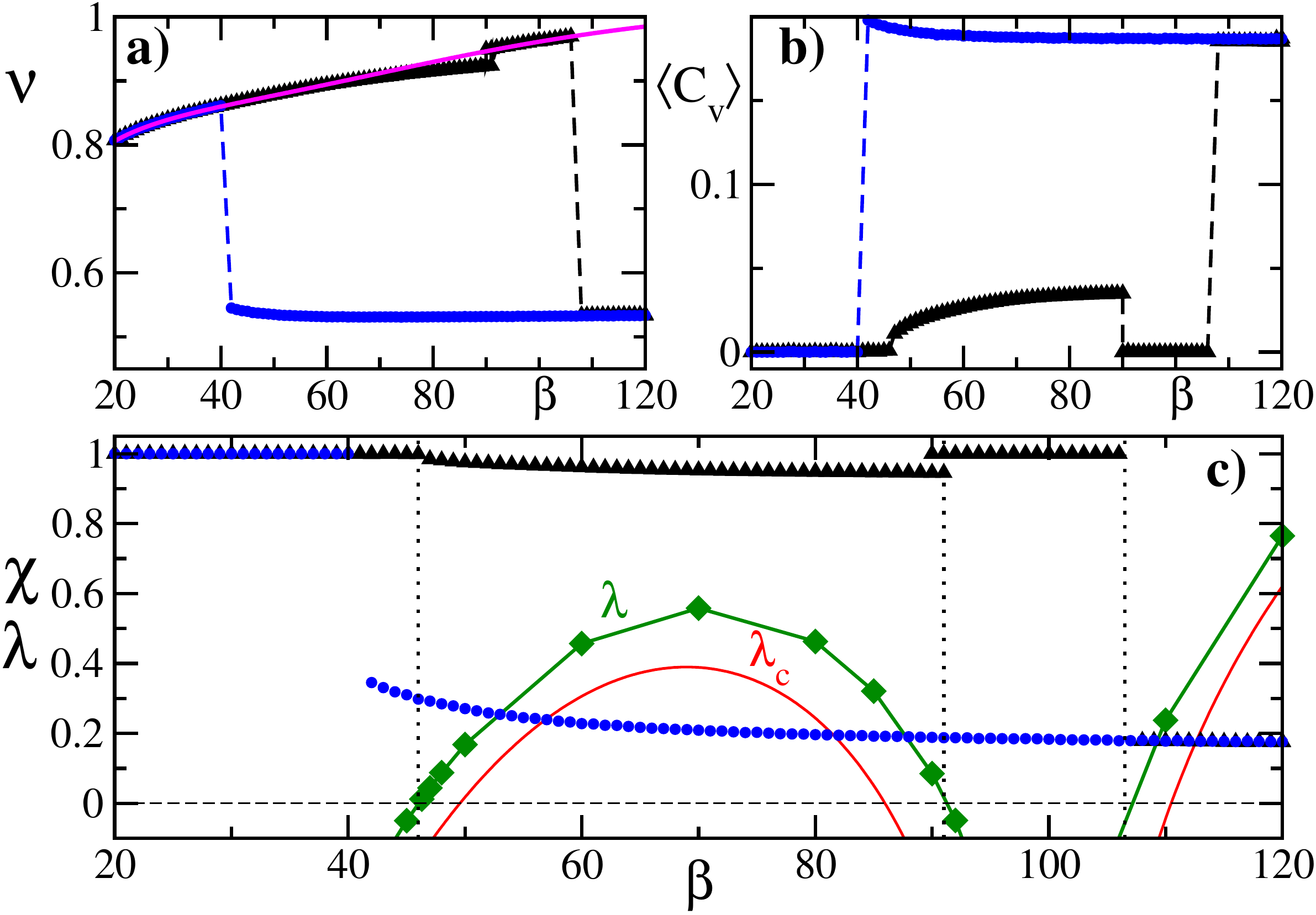}
\caption{The emergence of a bistable regime for nonidentical finite pulse-widths and PRC\textsubscript{1}. The parameter set is the same as in Fig.~\ref{fig:fig5}A with $N=10000$. Panels: a) mean firing rate $\nu$, b) mean coefficient of variations $\langle{C}_v\rangle$, and c) order parameter $\chi$ versus $\beta$. The blue circles and black triangles in all panels correspond to 
different initial conditions: fully random (circles), restricted to a tiny interval (triangles).
The narrow ICs are chosen to be in the order of $\delta_p=10^{-3}$. The green diamonds corresponds to the maximal Lyapunov exponent, and the red one is the conditional Lyapunov exponent as function of $\beta$. The magenta line (a) represents the semi analytic firing rate given in~Eq.\ref{eq:firingrate}. The horizontal dashed line (c) is a reference point ($\lambda=0$) in which the synchronous state changes its stability.}
\label{fig:fig9}
\end{figure*}

So far, the initial condition has been chosen by selecting independent, identically uniformly distributed random phases
and zero fields. Since it is known that discontinuous transitions are often accompanied by hysteretic phenomena,
we now explore this possibility. 
We start fixing a different type of initial conditions:
the phases are selected within a small interval of width $\delta_p$ (while the fields are set equal to 
zero and $\delta_t=10^{-4}$).~\footnote{In the simulations, it is crucial to set the time-step $\delta_t$ at least ten times smaller than $\delta_p$, in order to ensure that the spike times are properly handled during the integration process.}
Fig.~\ref{fig:fig9} combines the scenario presented in the previous section for a network with
$N=10000$ neurons and $\mu=0.3$ (the blue dots correspond to the content of Fig.~\ref{fig:fig5}A), 
with the results of the new simulations obtained for $\delta_p=10^{-3}$ (see black curves and triangles).
For $\beta \in I_1= [40,106]$, there is a clear bistability:
the new simulations reveal that $\chi \approx 1$, much above the typical CID value.

More precisely, $\chi < 1$ for $\beta\in I_2 \approx [46,91]$, while $\chi = 1$ for $\beta\in (I_1-I_2)$.
Since $\chi=1$ is accompanied by a vanishing $\langle C_v \rangle$, it is straightforward to conclude that this regime
is the periodic synchronous state, whose linear stability can be assessed quantitatively.

The conditional Lyapunov exponent $\lambda_c$ provides a semi-analytical approximate formula.
In Appendix~\ref{meanfield} we have derived Eq.~(\ref{eq:lyapcond}), whose implementation leads to the
red curve presented in Fig.~\ref{fig:fig9}(c). It provides a qualitative justification of the phase diagram:
for instance, we see that the synchronous solution is unstable in the interval $I_2$, where $\chi<1$.
By following the approach developed in Ref.~[\citen{af_eu_ap2020}], we can compute also
the maximal Lyapunov exponent $\lambda$: it is given by the maximal eigenvalue of a suitable random matrix.
The resulting values correspond to the green curve. The changes of sign of $\lambda$ coincide almost exactly with
the border of the intervals where the synchronous state ceases to be observed.

\begin{figure}
\centering
\includegraphics[width=.47\textwidth]{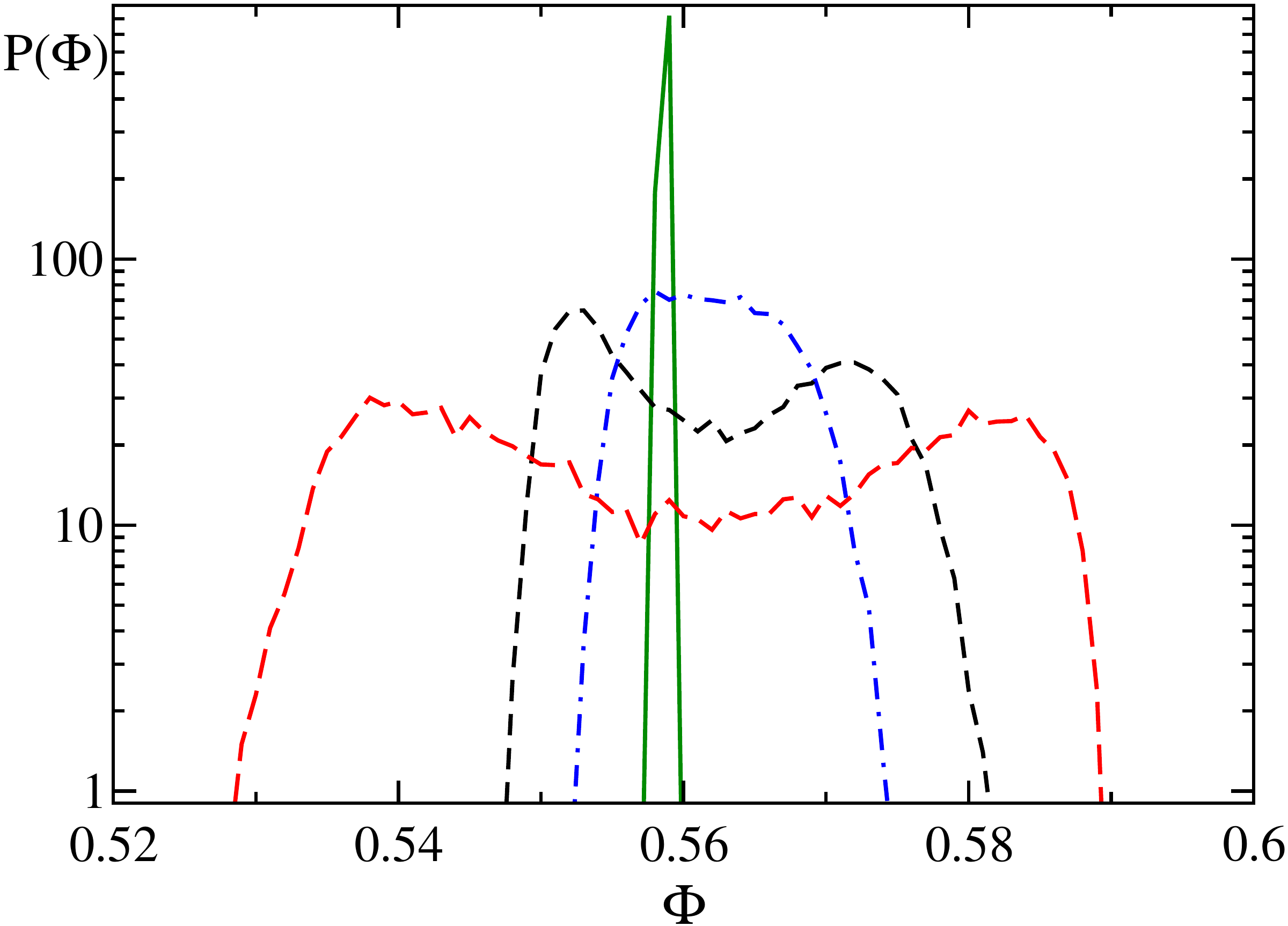}
\caption{Instantaneous probability distribution P($\Phi$) when centered around the same angle for different $\beta$ values:
70 (long-dashed red), 50 (short-dashed black), 48 (dot-dashed blue), and 46.3 (solid green). 
All snapshots correspond to the black triangles in Fig.~\ref{fig:fig9}.}
\label{fig:fig10}
\end{figure}

What is left to be understood is the regime observed within the interval $I_2$: it differs from the perfectly
synchronous state, but it is nevertheless nearly-synchronous. While approaching the left border of $I_2$,
where the synchronous state becomes stable, the width of the phase distribution progressively shrinks. This is
clearly seen in Fig.~\ref{fig:fig10}, where four instantaneous phase distributions are plotted for decreasing
$\beta$ values (from red to green curve).
The transition scenario occurring at the other edge of the interval $I_2$ appears to be different and further 
studies would be required. However, a comparative analysis of different models suggest that this regime follows
from a suitable combination of refractoriness and the shape of the PRC. As we suspect not to be very general,
we do not investigate it in further detail.

\begin{figure}
\centering
\includegraphics[width=.47\textwidth]{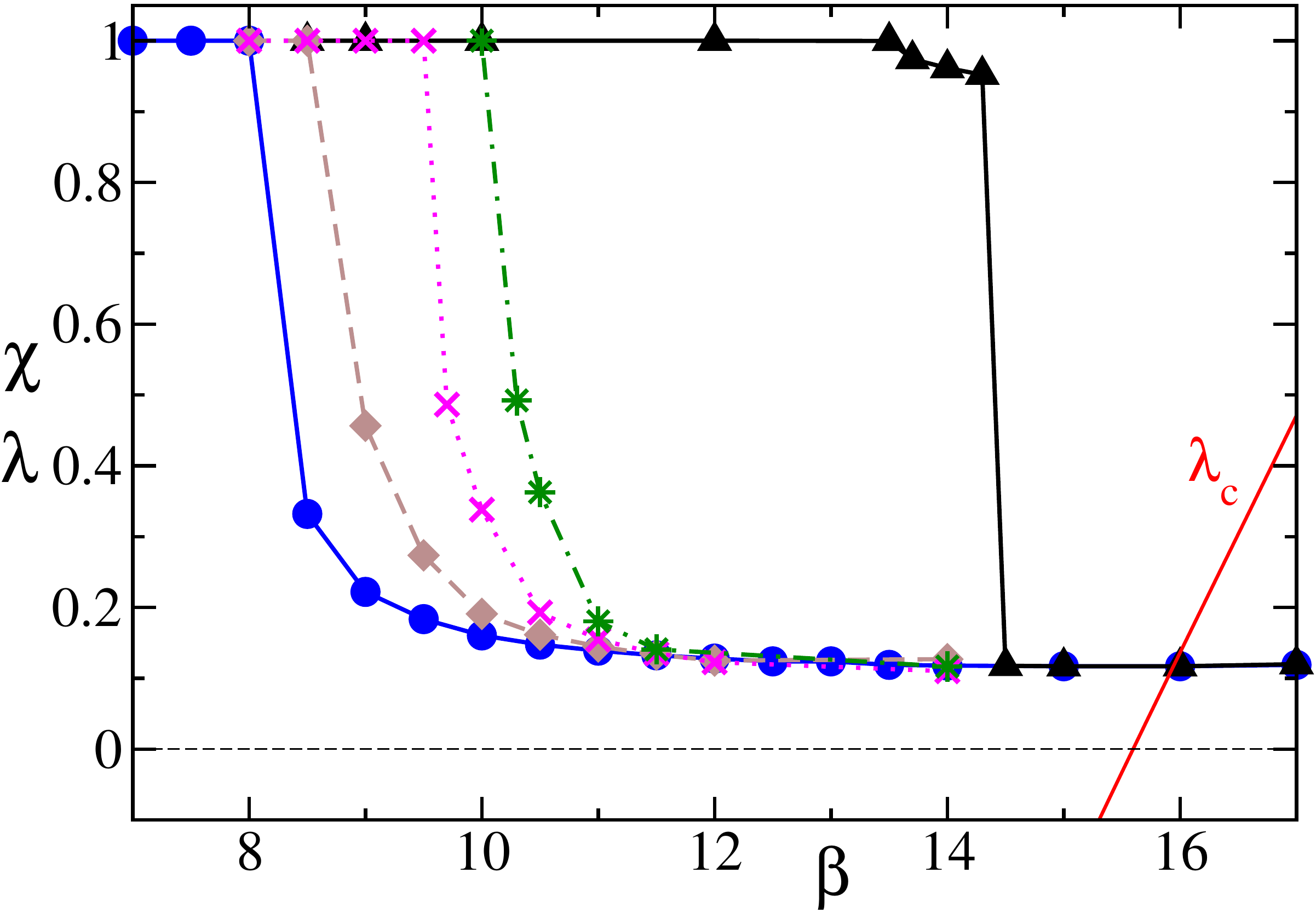}
\caption{Characterization of the global network dynamics for long finite-pulse widths, PRC\textsubscript{1}, $\mu=0.3$, 
and $\alpha=12$. The blue circles ($N=10000$), brown diamonds ($N=20000$), magenta cross ($N=40000$), and green stars ($N=80000$) correspond to full-range random ICs. The black triangles ($N=10000$) correspond to narrow ICs with $\delta_p=10^{-3}$. The red curve is the conditional Lyapunov exponent.}
\label{fig:fig11}
\end{figure}

Finally, we have considered broader pulses, to test the robustness of our findings.
More precisely, now we assume the pulse-widths $\alpha^{-1}$, $\beta^{-1}$ to be longer than the refractory time $t_r$ 
as observed in real neurons~\cite{ostojic2014,Gerstner2014}.
The results are displayed in Fig.~\ref{fig:fig11} for $\alpha=12$ and $\mu=0.3$. Once again, we see that CID extends to the region
where $\beta < \alpha$ and that the transition point moves progressively towards $\beta=\alpha$ upon increasing the network size
(see the different curves). On the other hand the strength of CID is significantly low ($\chi = 0.11$), 
possibly due to the relative smallness of the coupling strength.
Furthermore, the evolution of quasi-synchronous solutions ($\delta_p=10^{-3}$), reveals again bistability in
a relatively wide interval of $\beta$-values, $\beta\simeq 8.5-14.3$, which now extends beyond $\beta=\alpha$: a result,
compatible with the transversal stability (see the red curve for $\lambda_c$ in Fig.~\ref{fig:fig11}).

\subsection{Robustness}
\begin{figure*}
\centering
\includegraphics[width=.73\textwidth]{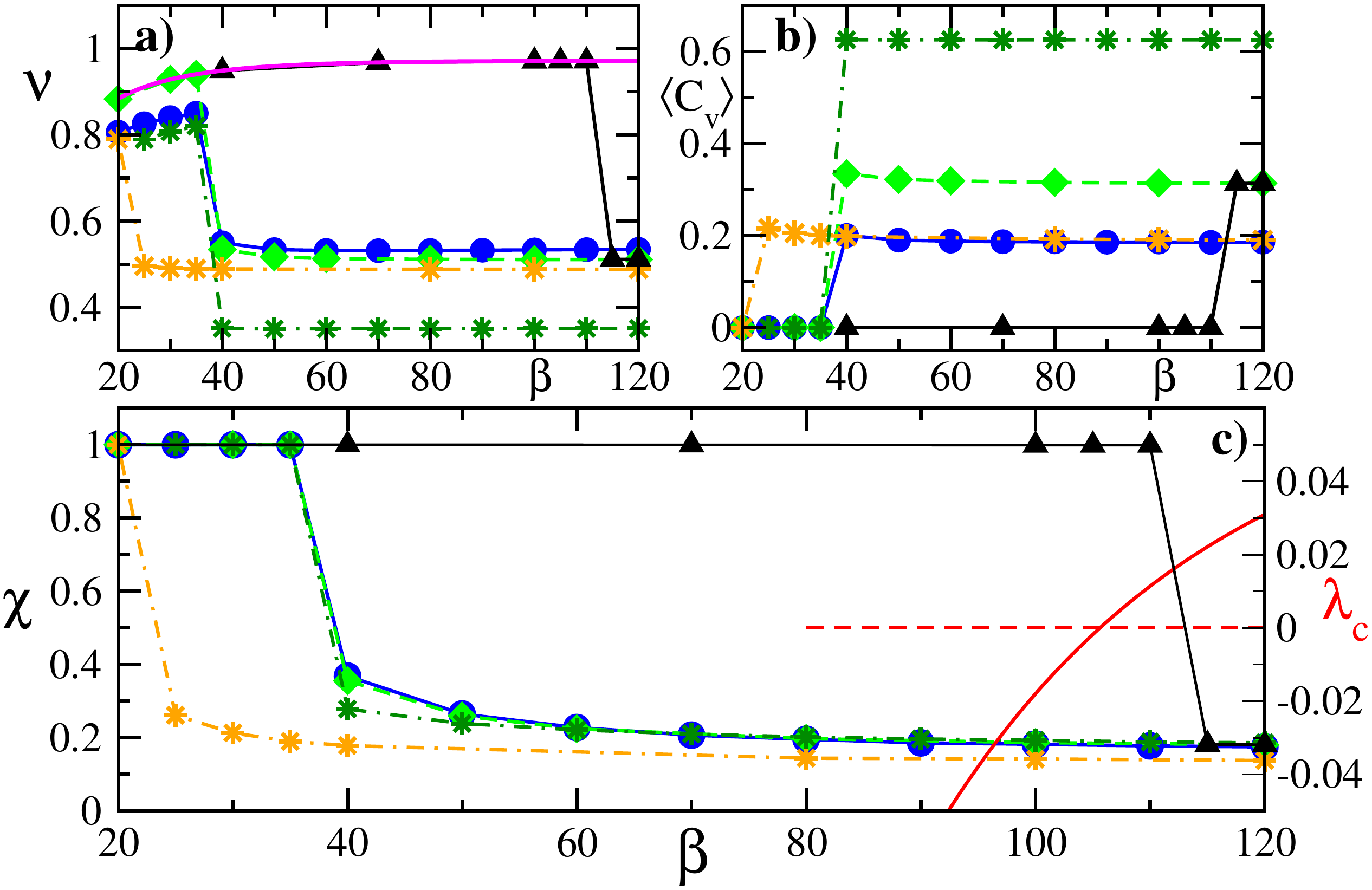}
\caption{The robustness for other PRCs.
The mean firing rate $\nu$, the mean coefficient of variations $\langle{C}_v\rangle$, and the order parameter $\chi$ vs. inhibitory pulse widths $\beta$ are shown in panel a), b) and c), respectively for $N=10000$ and $\alpha=100$. 
PRC\textsubscript{1} with random ICs is shown for $\mu=0.3$ (blue circles) as reference to Figs.~\ref{fig:fig5}(A) and \ref{fig:fig9}.
The PRC\textsubscript{2} with random ICs is depicted for $\mu=0.3$ (orange stars) and $\mu=0.7$ (green stars). Green diamond and black triangles result from PRC\textsubscript{3} and $\mu=0.3$. The former has been created with random ICs and the latter with strongly restricted ICs with $\delta_p=10^{-3}$  within the narrow basin of attraction for the synchronous attractor. The magenta curve (panel a) represents the semi-analytic firing rate for PRC\textsubscript{3} according to Eq.~\ref{eq:firingrate}.
Panel c) shows on the alternative y-axis also the conditional Lyapunov exponent $\lambda_c$ (red curve) for synchronous solutions and PRC\textsubscript{3}. The horizontal red dashed line is the null line to the axis on the right.
}
\label{fig:fig12}
\end{figure*}

In the previous sections we have investigated the dependence of CID on the spike-width as well as on the coupling strength.
Now, we examine the role of the PRC shape.
Following Fig.~\ref{fig:fig1}, we consider a couple of smoothened versions of PRC\textsubscript{1}, defined in section~\ref{sec:model}. 
The results obtained for a network of $N=10000$ neurons are reported in Fig.~\ref{fig:fig12}. 

All simulations have been performed for $\alpha=100$, while $\beta$ has been again varied in the range $[20,120]$.
In each panel, blue circles, orange stars and green diamonds have been obtained by setting $\mu=0.3$; they
correspond to PRC\textsubscript{1,2,3} respectively.
As a first general remark, the overall scenario is not strongly affected by the specific shape of the PRC.
The mean firing rate is approximately the same in all cases, while the coefficient of variation is substantially 
higher for the sinusoidal (and more realistic) PRC\textsubscript{3}.
Moreover, the order parameter for PRC\textsubscript{3} is remarkably close to that for PRC\textsubscript{1} (see panel c).

The most substantial difference concerns the transition from synchrony to CID, which occurs much earlier in PRC\textsubscript{2}.
On the other hand, the $\chi$-behavior of PRC\textsubscript{2} can be brought to a much closer agreement by increasing
the coupling strength (the green asterisks in Fig.~\ref{fig:fig12} refer to $\mu=0.7$).
This observation raises the issue of quantifying the effective amplitude of the coupling: PRCs are introduced
in Sec.~\ref{sec:model} are all functions whose maximum value is equal to 1. This does not exclude that the
effective amplitude may be significantly different, deviation that can be partially removed by
adjusting the value of the coupling constant $\mu$ as shown in Fig.~\ref{fig:fig12}.

Anyhow, these qualitative arguments need a more solid justification. In fact, 
in this last case (PRC\textsubscript{2} and $\mu=0.7$) $\langle{C}_v\rangle$
is significantly larger (above 0.6 instead of below 0.2),
consistently with the analysis carried out in Ref.~[\citen{eu_ap_at20}], where it is shown that a large coupling
strength induces a bursting phenomena in LIF neurons.

Finally, we investigate the presence of hysteresis in the case of PRC\textsubscript{3}.
The results, obtained by setting all parameters as in the previous cases,  are reported in Fig.~\ref{fig:fig12} (see
black triangles): they have been obtained by setting the initial spread of phases $\delta_p=10^{-3}$. 
Once again, there exists a wide parameter range where CID coexists with a stable synchronous regime.

At variance with the previous case (see Fig.~\ref{fig:fig9}), the synchronous state is always stable over the range $\beta\leq 110$. 
This is consistent with the variation of the conditional Lyapunov exponent, which 
does not exhibit an ``instability island".
As from Eq.~(\ref{eq:lyapcond}), $\lambda_c$ is the sum of two terms. In the  case of PRC\textsubscript{3}, 
the second one is absent because the PRC amplitude is zero at the reset value $\Phi_r=0$. 

\section{Conclusion and open problems}
In this paper we have discussed the impact of finite pulse-widths on the dynamics of a
weakly inhibitory neuronal network, mostly with reference to the sustainment and stability of the balanced regime.

In computational neuroscience, both exponential~\cite{Tsodyks2000} and $\alpha$-pulses~\cite{Olmi2012,Boari2019}
are typically studied. The former ones are simpler to handle, as they require one variable per neuron per field type
(inhibitory/excitatory); the latter ones, being continuous, are more realistic, but require twice as many variables.
In this paper we have selected exponential pulses to minimize the additional computational complexity.
We have prioritized the analysis of short pulses (about hundredth of the interspike interval) in order to single
out deviations from $\delta$-spikes. However tests performed for relatively longer spikes suggest that the
general scenario is substantially confirmed for ten-times longer pulses (a value compatible with the time scales
of AMPA receptors~\footnotetext[0]{The much slower NMDA receptors fall within another class of systems,
where a mean-field treatment is more appropriate.}\cite{Wang2010,Note0}).
The main changes observed when decreasing $\alpha$ down to 12 (starting from our reference 100)
is the disappearance of the quasi-synchronous regime for a small degree of asymmetry: this happens around
$\alpha \approx 60\sim70$.

Besides pulsewidth, the asymmetry between excitatory and inhibitory spikes
(a parameter which does not make sense in the case of $\delta$-pulses) plays a crucial 
role in the preservation of the balance between excitation and inhibition.
In fact, upon changing the ratio between excitatory and inhibitory pulse-width different regimes may arise.
The role of time scales is particularly evident in the synchronous regime, where the overall field is the superposition of
two suitably weighted exponential shapes with opposite sign: depending on the time of observation, 
the effective field may change sign signalling a prevalence of either inhibition or excitation.

Altogether CID is robust when inhibitory pulses are shorter than excitatory ones
(this is confirmed by the corresponding instability of the synchronous regime).
More intriguing is the scenario observed in the opposite case, when CID and synchrony maybe simultaneously
stable.
A finite-size analysis performed by simulating increasingly large networks shows that the hysteretic
region progressively shrinks, although it is still prominent - especially for weak coupling - for
$N=80000$, when the connectivity of our networks ($K=8000$) is comparable to that of the mammalian brain.
Anyhow, on a purely mathematical level, one can argue that the transition from CID to synchrony 
eventually occurs for identical widths.

Further studies are definitely required to reconstruct the general scenario, since the dynamics depends on several 
parameters.
Here, we have explored in a preliminary way the role of the PRC shape:
so long as it is almost of Type I, the overall scenario is robust. 

Finally, the transition from CID to synchrony requires more indepth studies.
A possible strategy consists in mimicking the background activity as a pseudo-stochastic
process, thereby writing a suitable Fokker-Planck equation. However, at variance with the 
$\delta$-spike case, here additional variables would be required to account for the dynamics of the inhibitory/excitatory
fields.

\begin{acknowledgments}
Afifurrahman was supported by the Ministry of Finance of the Republic of Indonesia through the Indonesia Endowment Fund for Education (LPDP) (grant number: PRJ-2823/LPDP/2015).
\end{acknowledgments}

\begin{data availability}
The data that support the findings of this study are available from the corresponding author upon reasonable request.
\end{data availability}

\appendix

\section{\label{meanfield}Mean field model for finite-width pulse}
We investigate the stability of the period-1 synchronous state through the conditional Lyapunov exponent. 
This regime is characterised by a synchronous threshold-passing of all oscillators 
leading to exactly the same exponentially decaying excitatory and inhibitory field for all oscillators.
The synchronous solution $\Phi(t)$ with a period $T$ of Eqs.~(\ref{eq:mod1},\ref{eq:mod2}) is obtained by integrating the equation
\begin{align}
\begin{cases}
   \dot{\Phi} &= 1 + \frac{\mu}{\sqrt{K}}\Gamma{(\Phi})(E(t)-I(t))\label{eq:p1} \\
   E(t) &= E_{\circ}\text{e}^{-\alpha t} \\
   I(t) &= I_{\circ}\text{e}^{-\beta t}
\end{cases}   
\end{align}
where $$E_{\circ}=\frac{K_e \alpha}{1-\text{e}^{-\alpha T}}, \hspace{1cm} I_{\circ}=\frac{gK_i \beta}{1-\text{e}^{-\beta T}}.$$

The fields follow an exponential decay with the initial amplitudes $E_{\circ},I_{\circ}$ for the excitatory and inhibitory field, respectively. In order to determine the stability of the synchronous state, we first need to find the period $T$ via a self-consistent iterative approach. Setting the origin $t=0$ as the time when the phase is reset to zero, we define $T$ 
as a time when the phase variable reaches its maximal value i.e. $\Phi(T)=1$. We integrate the phase starting from $\Phi=0$ up to $\Phi(T)$ by initially imposing arbitrary non-zero values for $E_{\circ}$ and $I_{\circ}$. 
The procedure is then repeated with updated values of the initial field amplitudes $E_{\circ},I_{\circ}$, until
convergence to a fixed point is attained. 
The firing rate is given by,
\begin{equation}\label{eq:firingrate}
\tilde{\nu}\equiv\frac{1}{T}
\end{equation}

The conditional (also known as transversal) Lyapunov exponent is a simple tool to assess the stability of the synchronous regime. 
It quantifies the stability of a single neuron subject to the external periodic modulation 
resulting from the network activity. 
The transversal Lyapunov exponent is the growth rate $\lambda_c$ of an infinitesimal perturbation.
Let us denote with $\delta t_r$ the time shift at the end of a refractory period. The corresponding phase shift 
is \cite{af_eu_ap2020}
\begin{equation}\label{eq:transv0}
\delta\phi_r = \dot \Phi(t_r)\delta t_r = \left \{1+ \frac{\mu}{\sqrt{K}} \Gamma(0) [E(t_r)-I(t_r)]\right \}\delta t_r \; .
\end{equation}
From time $t_r$ up to $t_m$ the phase shift evolves according to,  
\begin{equation}\label{eq:transv}
\dot{\delta\phi} = \frac{\mu}{\sqrt{K}}\Gamma'(\Phi) (E(t)-I(t))\delta\phi \; ,
\end{equation}
where $t_m$ is the minimum between the time when PRC\textsubscript{1} drops to zero 
and the time when the threshold is reached
(in either case, we neglect the variation of field dynamics, since the field is treated as an external forcing).
As a result,
\begin{equation}\label{eq:tr1}
    \delta\phi = \mathrm{e}^D \delta\phi_r \; ,
\end{equation}
where, 
\begin{equation}\label{eq:tr2}
D=\int_{t_r}^{t_m} \frac{\mu}{\sqrt{K}}\Gamma'(\Phi)(E(t)-I(t)) dt.
\end{equation}

The corresponding time shift is
\[
\delta t_s = \frac{\delta\phi}{\dot {\Phi} (t_m) }
\]
where $\dot {\Phi} (t_m)$ is the velocity at $t_m$.  The shift $\delta t_s$ carries over unchanged until first the threshold $\Phi_{th}=1$ is crossed and then 
the new refractory period ends. Accordingly, from Eqs.~(\ref{eq:transv0},\ref{eq:tr1}),
the expansion $R$ of the time shift over one period (a sort of Floquet multiplier) can be written as

\begin{equation}\label{eq:R}
R = \frac{\delta t_s}{\delta t_r} = 
\frac{1+ \frac{\mu}{\sqrt{K}} \Gamma(0) [E(t_r)-I(t_r)]}{\dot {\Phi} (t_m)} \mathrm{e}^D
\end{equation}
This formula is substantially equivalent to Eq.~(54) of Ref.~[\citen{Olmi2014}] ($\Lambda_{ii}$ corresponds to $R$),
obtained while studying a single population under the action of $\alpha$-pulses. 
An additional marginal difference is that while in Ref.~[\citen{Olmi2014}] the single neuron dynamics is described by a non uniform
velocity field $F(x)$ and homogeneous coupling strength, here we refer to a constant velocity and a phase-dependent
PRC, $\Gamma(\Phi)$. 

The corresponding conditional Lyapunov exponent is
\begin{equation}\label{eq:lyapcond}
\lambda_c = \frac{\ln |R| }{T}=\frac{D+\ln{{\left|[1+\frac{\mu}{\sqrt{K}}\Gamma(0)(E(t_{r})-I(t_r))]/{\dot {\Phi} (t_m)}\right|}}}{T}.
\end{equation}
The formula~(\ref{eq:lyapcond}) is valid for all PRCs as long as $t_m$ is replaced by $T$.
The formula~(\ref{eq:lyapcond}) is the sum of two contributions: the former one accounts for the linear stability 
of the phase evolution from reset to threshold ($D/T$); the latter term arises from the different velocity
(frequency) exhibited at threshold and at the end of the refractory period.
Notice that in the limit of short pulses, the field amplitude at time $t_m$ is effectively negligible, and
one can thereby
neglect the effect of the fields and assume $\dot {\Phi} (t_m)= 1$.

In mean-field models, the conditional Lyapunov exponent coincides with the exponent obtained by implementing
a rigorous theory which takes into account mutual coupling. 


%

\end{document}